\documentclass{ws-mpla}

\def\p  {\pi}

\def\vph {\varphi}

\def\cA {\cal A}

\def\be {\begin{equation}}
\def\ee {\end{equation}}
\def\beq{\begin{equation}}
\def\eeq{\end{equation}}
\def\bea {\begin{eqnarray}}
\def\eea {\end{eqnarray}}
\def\br{\begin{eqnarray}}
\def\er{\end{eqnarray}}
\def\bc {\begin{center}}
\def\ec {\end{center}}
\def\bi {\begin{itemize}}
\def\ei {\end{itemize}}
\def\benu{\begin{enumerate}}
\def\eenu{\end{enumerate}}
\newcommand{\bdm}{\begin{displaymath}}
\newcommand{\edm}{\end{displaymath}}

\def\la {\label}
\def\le {\left}
\def\ri {\right}
\def\l{\left}
\def\r{\right}
\def\pa {\partial}

\def\fr {\frac}

\def\laq{\hbox{~}\raise 0.4ex\hbox{$<$}\kern -0.8em\lower 0.62ex\hbox{$\sim$}\hbox{~}}
\def\gaq{\hbox{~}\raise 0.4ex\hbox{$>$}\kern -0.7em\lower
 0.62ex\hbox{$\sim$}\hbox{~}}

\newcommand{\SEN}{S_{_{\rm ent}}}

\newcommand{\ket}[1]{\mbox{$\mid #1\,\rangle$}}

\newcommand{\lPl}{\ell_{_{\rm Pl}}}
\newcommand{\SBH}{S_{_{\rm BH}}}

\newcommand{\AHo}{{\mathcal A}_{_{\rm H}}}
\newcommand{\rHo}{r_{_{\rm H}}}

\begin{document}
\markboth{Shankaranarayanan}{Do subleading corrections to Bekenstein-Hawking entropy hold
the key to quantum gravity?}
\title{Do subleading corrections to Bekenstein-Hawking entropy hold
the key to quantum gravity?}
\author{\footnotesize S. Shankaranarayanan}
\address{
Institute of Cosmology and Gravitation, 
University of Portsmouth, \\
Mercantile House, Portsmouth P01 2EG, U.K.
\footnote{\tt Email: shanki.subramaniam@port.ac.uk}}
\maketitle

\begin{abstract}
Black-holes are considered to be theoretical laboratories for testing
models of quantum gravity. It is usually believed that any candidate
for quantum gravity must explain the microscopic origin of the
Bekenstein-Hawking ($S_{_{\rm BH}}$) entropy. In this letter, we argue
(i) the requirement for a candidate approach to go beyond $S_{_{\rm
BH}}$ and provide generic subleading corrections, and (ii) the
importance to {\it disentangle} and identify the degrees of freedom
leading to $S_{_{\rm BH}}$ and its subleading corrections. Using the
approach of entanglement of modes across the horizon, we show that the
microscopic degrees of freedom that lead to $S_{_{\rm BH}}$ and
subleading corrections are different. We further show, using
microcanonical and canonical ensemble approaches, that the quantum
entanglement predicts generic power-law corrections to $S_{_{\rm BH}}$
and that the corrections can be identified with the kinematical
properties of the event-horizon.
\end{abstract}

\ccode{PACS nos: 04.60.-m, 04.70.-s, 04.70.Dy, 03.65.Ud} 

In any physical theory, entropy takes an unique position among other
physical quantities. This is due to the fact that the entropy relates
the macroscopic and microscopic structure of a system  through the Boltzmann 
relation $S = k_B \, \ln \Omega$ (where $\Omega$ is the total number 
of accessible states) \cite{Wehrl:1978}.  The entropy of black-hole is 
unique and distinct from that of other physical systems and, hence, 
is not a surprise that entropy has taken a pivotal role in understanding 
black-hole properties:\\
(i) Black-hole entropy is not extensive unlike, for instance, the
entropy of an ideal gas. Its leading order ---
Bekenstein-Hawking entropy $S_{_{\rm BH}}$ --- is proportional to the
area of the event horizon ($\cA_{\rm H}$) of the black-hole
\cite{Bekenstein:1972c,Bekenstein:1973b} i. e.
\begin{equation} \la{eq:SBH} 
S_{_{\rm BH}} = \le(\frac{k_{_{B}}}{4}\ri)
\frac{\cA_{\rm H}}{\ell_{_{\rm Pl}}^2} \quad 
\mbox{where} \quad {\ell_{_{\rm Pl}}} \equiv \sqrt{\frac{G\hbar}{c^3}} 
\mbox{~~~is the Planck length} 
\, .
\end{equation}
Unlike ideal gas, the finiteness of the black-hole entropy 
requires that the matter and(or) gravity have quantum description which is 
evident from Eq.~(\ref{eq:SBH}). \\
(ii) It is still unclear, what are {\it the} microscopic degrees of
freedom (DOF) leading to black-hole entropy?  Currently, there are
several approaches starting from counting states (by assuming
fundamental structures) \cite{Strominger-Vafa:1996a,Ashtekar-Baez:1997a,Carlip:2002a} 
to Noether charge \cite{'tHooft:1984a,Bombelli-Koul:1986,Srednicki:1993,Wald:1993a,Das-Shanki:2005a}. 
Although, none of these approaches can be considered to be complete; all of them --- within
their domains of applicability --- by counting certain microscopic
states yield (\ref{eq:SBH}).  This is in complete contrast to other
physical systems, such as ideal gas, where quantum DOF are uniquely
identified and lead to the classical thermodynamic entropy. This is
one of the few areas of physics where the semi-classical result
dictates and, help to, identify the quantum theory.

The above discussion raises an important question: {\it Is it
sufficient for an approach to reproduce (\ref{eq:SBH}) or need to go
beyond the Bekenstein-Hawking entropy?}  As we know, $S_{_{\rm BH}}$
is a semi-classical result and there are strong indications that
Eq. (\ref{eq:SBH}) is valid for large black holes [i.e. $\cA_{\rm H}
\gg \ell_{_{\rm Pl}}$]. However, it is not clear, whether this
relation will continue to hold for the Planck-size
black-holes. Besides, there is no reason to expect that $\SBH$ to be
the whole answer for a correct theory of quantum gravity. In order to
have a better understanding of black-hole entropy, it is imperative for
any approach to go beyond $\SBH$ and identify the subleading
corrections.

This raises a related question: {\it Are the quantum DOF that
contribute to $\SBH$ and its subleading corrections, identical or
different?} In general, the quantum DOF can be different. However,
several approaches in the literature \cite{correction1,correction2,correction3,correction4} 
that do lead to subleading corrections either assume that the quantum DOF are
identical or do not {\it disentangle} DOF contribution to $S_{_{\rm
BH}}$ and the subleading corrections.

In this letter, we show that the quantum DOF that contribute to the
Bekenstein-Hawking entropy and its subleading corrections are
different. Using the approach of entanglement, we show that it is
possible to {\it disentangle DOF contributions}. To isolate different
contributions and elucidate their role in black-hole entropy, we
obtain entanglement entropy ($\SEN$) in two different statistical ---
microcanonical and canonical --- ensembles. Using the Schr\"odinger
representation of the quantum fields, we show that entanglement
predicts generic power-law corrections to $S_{_{\rm BH}}$.

So, what is entanglement (entropy) and how can it possibly be the
source of black-hole entropy? Given a joint quantum system $\{A B\}$,
entanglement refers to the quantum correlation between the sub-systems
$A$ and $B$. Entanglement is quantified by the entropy, $\SEN$, of the
reduced density matrix $\rho_{\alpha}$ of either of the subsystems
defined as
\begin{equation}
\label{eq:VNS}
\SEN = - {\rm Tr}[\rho_\alpha \ln(\rho_\alpha)] 
\qquad \qquad \alpha \in \l\{A, B\r\} \, .
\end{equation}
The relation between $\SEN$ and black-hole entropy can be understood
from the fact that both are (i) quantum effects with no classical
analogues and (ii) associated with the existence of horizon
\cite{Das2007c}.

Let us now go to the details and see how using entanglement we can
identify DOF contributing to $S_{_{\rm BH}}$ and the power-law
corrections. We consider a massless scalar field $(\varphi)$
propagating in an asymptotically flat, four-dimensional black-hole
background given by the Lema\^itre line-element\footnote{The
motivation for the choice of scalar fields is given in
Ref. \cite{Das2007c}.}:
\begin{equation}
\label{eq:Lema}
ds^2 
= - d\tau^2 + [1 - f(r)] d\xi^2 + r^2 \l[d\theta^2 + 
\sin^2\theta \, d\phi^2\r] \, ,
\end{equation}
where $r$ is the radial coordinate in the Schwarzschild coordinate
system and is related to $(\xi,\tau)$ by the relation $\xi - \tau =
\int dr/{\sqrt{1 - f(r)}}$. The Hamiltonian of the scalar field propagating 
in the above line-element is 
\begin{equation}
H(\tau) = \frac{1}{2} \int_{\tau}^{\infty} \!\!\!\!
d\xi \le[ \frac{1}{r^2 \sqrt{1 - f(r)}} \Pi_{_{lm}}^2 
+ \frac{r^{2}}{\sqrt{1 - f(r)}} \le(\pa_{\xi} \vph_{_{lm}}\ri)^2
 + l(l + 1)\sqrt{1 - f(r)} \, \vph_{_{lm}}^2 \ri] \,  ,
\label{eq:ham1}
\end{equation}
where $\varphi_{lm}$ is the spherical decomposed field and $\Pi_{lm}$
is the canonical conjugate of $\varphi_{lm}$. [For simplicity of the
notation, we will suppress the subscripts $(lm)$.] Although, the above
Hamiltonian is time-dependent, there are several advantages of
Lema\^itre coordinate over the Schwarzschild coordinate: (i) the
former is not singular at the horizon ($\rHo$) as opposed to the
latter, and (ii) $\xi$ (or $\tau$) are space(or, time)-like everywhere
while $r$ is space-like only for $r > \rHo$.

Having obtained the Hamiltonian, the next step is quantization. We use
Schr\"odinger representation since it provides a simple and intuitive
description of vacuum states for time-dependent Hamiltonian \cite{Hatfield:1992-bk}.
Formally, we take the basis vector of the state vector space to be the
eigenstate of the field operator ${\hat \varphi}(\tau, \xi)$ on a
fixed $\tau$ hypersurface, with eigenvalues $\varphi(\xi)$ i. e.
${\hat \varphi}(\tau,\xi)\ket{\varphi(\xi),\tau} =
\varphi(\xi)\ket{\varphi(\xi),\tau}$.
The quantum states are explicit functions of time and are represented
by wave functionals $\Psi[\varphi(\xi),\tau]$ which satisfy the
functional Schr\"odinger equation:
\begin{equation}
\label{schrod}
i \frac{\partial \Psi}{\partial \tau} = \int_{\tau}^{\infty} \!\!\!
d\xi  \, H(\tau) \, \Psi[\varphi(\xi),\tau] \, .
\end{equation}

To proceed with the evaluation of $\SEN$: \\ 
(i) We assume that the Hamiltonian evolves adiabatically. Technically,
this implies that the evolution of the late-time modes leading to
Hawking particles are negligible. In the microcanonical ensemble
[where the total energy is fixed], this assumption translates to the
weak time-dependence of the functional ($\Psi[\varphi(\xi),
\tau]$). In the canonical ensemble [where the temperature is fixed],
this corresponds to black-hole in thermal equilibrium and
$\Psi[\varphi(\xi), \tau]$ is approximated as a WKB functional. \\
(ii) We then obtain $\rho_{\alpha}$ by tracing the region enclosing
the horizon [$\xi \to (\rHo, \infty)$] and use Eq. (\ref{eq:VNS}) to
determine $\SEN$\footnote{$\SEN$ is generally divergent in continuum
theories. Therefore usually we assume an ultraviolet cutoff $\lPl$ to
regulate the quantum field theory. Below we assume that this is just a
technical issue and that we can always have such a regularization see
Refs. \cite{Sarkar:2007a,Brout:2008qf} and references therein.}.

\noindent {\it Microcanonical ensemble:} The Hamiltonian 
(\ref{eq:ham1}) at a fixed Lema\^itre time $\tau = \tau_0 \equiv 0$
reduces to \cite{Das2007c}
\begin{equation}
H_{_{F}} = \fr 1 2 \int_0^\infty dr
\le\{\p^2(r) + r^2 \le[\fr{\pa}{\pa r} \le(\fr{\varphi
(r)}{r}\ri)\ri]^2 + \fr{l(l+1)}{r^2}~\varphi^2(r)\ri\} \, ,
\label{eq:ham2}
\end{equation}
where $\pi$ is a canonical transformed variable given by, $\Pi = r
\sqrt{1 - f(r)} \, \pi$. This is the Hamiltonian of a scalar field in
flat space-time and is valid for any fixed Lemaitre time. ${\hat
\varphi}$ are time-independent and $\Psi[\varphi]$ satisfies the
time-independent Scr\"odinger equation (\ref{schrod}).  For
simplicity, let us choose the wave-functional to be superposition
\cite{Das2007c} of the ground $\Psi_0[\varphi]$ and first-excited
state $\Psi_1[\varphi]$, i. e.,
\begin{equation}
\Psi[\varphi(r)] = c_0 \Psi_0[\varphi] + c_1 \Psi_1[\varphi] \, ,
\end{equation}
where $c_0, c_1$ are constants satisfying $|c_0|^2 + |c_1|^2 = 1$, and
\begin{equation}
\Psi_0[\varphi] =  \prod_k \l(\frac{\omega_k}{\pi}\r)^{1/4} 
\exp\l[-\frac{1}{16 \pi^3} \varphi^2(|k|)\r] ; 
\Psi_1[\varphi] = \l( \frac{2 \omega_{k_1}}{(2 \pi)^3}\r) \varphi(k_1) 
\Psi_0[\varphi] \, .
\end{equation}
Following the procedure discussed in the previous page, the numerical
evaluation of the density matrix leads to following best fit for the
microcanonical entanglement entropy \cite{Das2007c}
\begin{equation}
\SEN^{\rm mc} = S_{_{\rm BH}} \l[1 + 
a_1 \l(\frac{{\cA}_{\rm H}}{\ell_{_{\rm Pl}}^2}\r)^{-\nu}\r] 
\end{equation}
where $a_1 \propto |c_1|$ and $\nu > 0$. This is the first result of
this letter. It is instructive to stress the implications of the
result: \\
(i) $\SEN^{\rm mc}$ is obtained for a scalar field in a flat
space-time. Thus, $S_{_{\rm BH}}$ and the subleading corrections 
can be uniquely identified with the correlation of the 
quantum states. \\
(ii) For the pure vacuum wave-functional, $a_ 1
= 0$ and $\SEN^{\rm mc}$ is identical to Bekenstein-Hawking
entropy. This clearly shows that the entanglement entropy of 
ground state leads to the area law and the excited states contribute 
to the power-law corrections. \\
(iii) For large black-holes, power-law correction falls off rapidly
and we recover $\SBH$. However, for the small black-holes, the second
term dominates and black-hole entropy is no more proportional to area.
%
Physical interpretation of this result is immediately apparent. In the
large black-hole (or low-energy) limit, it is difficult to excite the
modes and hence, the ground state modes contribute significantly to
$\SEN$. However, in the small black-hole (or high-energy) limit,
larger number of field modes can be excited and hence they contribute
significantly to $\SEN$.

Having established that the quantum DOF that contribute to $\SBH$ are
different from the one that contribute to the corrections, our next
step is to identify the power-law contributions to the kinematical
properties of the horizon. For this we obtain $\SEN$ in the canonical
ensemble.

\noindent {\it Canonical ensemble:} In the adiabatic 
limit, the ansatz for the wave-functional is
\begin{equation}\label{vacans}
\Psi[\varphi(\xi), \tau] = P[\varphi] \exp\l[\frac{i}{\hbar} 
S[\varphi(\xi, \tau)] \r]
\end{equation}
where $S$ is the Hamilton-Jacobi corresponding to (\ref{eq:ham1}) and
$P[\varphi]$ is the 1-loop term.  Using the relation between $\rho$
and partition function $Z$,
\begin{equation}
\rho(\beta) = \frac{1}{2 \pi i} \int_{-i \infty}^{i \infty} 
Z(\beta) \, e^{\beta E} \, dE
\end{equation}
and following the procedure discussed earlier, we get \cite{Sarkar:2007a}
\begin{equation}
\SEN^{\rm c}
= \SBH + {\cal F} \; \log\l(\frac{\AHo}{\lPl^2} \r),  
\quad \mbox{where} \quad 
{\cal F} = - \frac{1}{60} \, f''(\rHo)\; \rHo^2 
- \frac{1}{10} \, \kappa\, \rHo\,,
\label{eq:4D-Tot-SBW}
\end{equation}
$\kappa$ is surface gravity and $f''(\rHo)$ is second derivative of
metric at the horizon. This is the second key result of this letter
regarding which we would like to stress a few points: \\
(i) This is a master equation and gives the entropy corresponding to a
general spherically symmetric black-hole space-time. The sub-leading
corrections depend only on the kinematical properties of black-hole
i. e. surface gravity and second deriviative of metric function. [It
should be noted that this form is unique for all orders of the WKB
approximation and does not depend on third and higher order
derivatives of the metric \cite{Sarkar:2007a}.] \\
(ii) ${\cal F}$ is a constant --- and hence, subleading corrections
are purely logarithmic --- only, if $\kappa \propto \rHo^{-1}$ and
$f''(\rHo) \propto \rHo^{-2}$. This uniquely corresponds to
Schwarzschild space-time.  For any other black-hole space-times like,
for instance, Reissner-N\"ordstrom, Schwarzschild-(Anti)de Sitter, we
have
\begin{equation}
\SEN^{\rm c} = \SBH - \frac{\pi^{1/2}}{15} \l(\frac{\lPl}{\AHo}\r)^{-1/2} 
\log\l(\frac{\AHo}{\lPl^2} \r) + \mbox{Higher contributions}
\end{equation}
(iii) As in the microcanonical ensemble, (a) in the large black-hole
limit the power-law corrections fall off rapidly and we recover
$S_{_{\rm BH}}$ (b) in the small black-hole limit, the second term
dominates and the black-hole entropy is not proportional to area.
 
In summary, we have emphasized the importance of the subleading
corrections to $S_{_{\rm BH}}$ and their role in identifying the
structure of the quantum theory. Using the approach of entanglement,
we have shown that the quantum DOF that lead to $S_{_{\rm BH}}$ {\it
need not necessarily contribute significantly} to subleading terms.
Since entanglement is a quantum effect and should be present in any
quantum theory, the results presented here do have implications beyond
the semiclassical regime and hence, the subleading corrections do seem
to hold the key to unlock the mysteries of quantum gravity.

The author wishes to thank Saurya Das, Hubert Lampeitl, 
Sudipta Sarkar, L. Sriramkumar and Sourav Sur for discussions. 
The work is supported by the Marie Curie Incoming International 
Grant IIF-2006-039205.


\end{document}